\documentclass[aps,prb,10pt,twocolumn,floatfix,showpacs,superscriptaddress]{revtex4-1}
\usepackage{graphicx,amsmath,amsfonts,bm}
\usepackage[colorlinks=true, citecolor=blue, linkcolor=blue, urlcolor=blue]{hyperref}
\usepackage[all]{hypcap}
\usepackage{color}

\begin{document}

\title{Decaying spectral oscillations in a Majorana wire with finite coherence length}
\author{C. Fleckenstein}
\email{christoph.fleckenstein@physik.uni-wuerzburg.de}
\affiliation{Institute of Theoretical Physics and Astrophysics, University of W\"urzburg, 97074 W\"urzburg, Germany}
\author{F. Dom\'{i}nguez}
\affiliation{Institute of Theoretical Physics and Astrophysics, University of W\"urzburg, 97074 W\"urzburg, Germany}
\author{N. Traverso Ziani}
\affiliation{Institute of Theoretical Physics and Astrophysics, University of W\"urzburg, 97074 W\"urzburg, Germany}
\author{B. Trauzettel}
\affiliation{Institute of Theoretical Physics and Astrophysics, University of W\"urzburg, 97074 W\"urzburg, Germany}
\begin{abstract}
Motivated by recent experiments, we investigate the excitation energy of a proximitized Rashba wire in the presence of a 
position dependent pairing. 
In particular, we focus on the spectroscopic pattern produced by the overlap between two Majorana bound states that appear for values of the Zeeman field smaller than the value necessary for reaching the bulk topological superconducting phase.
The two Majorana bound states can arise because locally the  wire is in the topological regime. 
We find three parameter ranges with different spectral properties: crossings, anticrossings and asymptotic reduction of the energy as a function of the applied Zeeman field. 
Interestingly, all these cases have already been observed experimentally. 
Moreover, since an increment of the magnetic field implies the increase of the distance between the Majorana bound states, the amplitude of the energy oscillations, when present, gets reduced. 
The existence of the different Majorana scenarios crucially relies on the fact that the two Majorana bound states have distinct $k$-space structures. We develop analytical models that clearly
explain the microscopic origin of the predicted behavior.

\end{abstract}

\maketitle

\section{Indroduction}
Majorana fermions are fermionic particles which are their own antiparticles, i.e.~$\gamma=\gamma^\dagger$.\cite{Majorana1937a}
In condensed matter physics, these particles arise as quasiparticle excitations in topological superconductors.\cite{Volovik1999a, Read2000a, Kitaev2001a}
Models for engineering topological superconductivity and its detection have been the matter of an 
extensive research over the last decade.
The common ingredient in most of these models consists in proximitizing s-wave superconductivity into a 
system with strong spin-orbit interaction.\cite{Fu2008a, Linder2010a, Lutchyn2010a,Oreg2010a,Choy2011a}
Their interest is not only fundamental but also practical because 
they exhibit non-abelian statistics\cite{Ivanov2001a,Nayak2008a, Alicea2011a} and therefore, can potentially be used in protocols for topological quantum computation.
Signatures of Majorana bound states (MBSs) are predicted to appear in electrical conductance,\cite{Bolech2007a, Law2009a, Wimmer2011a, Prada2012a}, thermal conductance,\cite{Wimmer2010a,Akhmerov2011a, Sothmann2016a} 
ac-Josephson effect,\cite{Kwon2004a,Jiang2011a, Badiane2011a, San-Jose2012a, Dominguez2012a, Houzet2013a, Pikulin2011b, Dominguez2017b, Pico2017a} and studying the skweness of the $4\pi$-periodic supercurrent.\cite{Tkachov2013a}
Indeed, experimental measurements confirm some of these predictions in the conductance,\cite{Mourik2012a, Rodrigo2012a, Deng2012a, Das2012a, Churchill2013a}  
Shapiro steps,\cite{Rokhinson2012a, Wiedenmann2016a,Bocquillon2016a}, Josephson radiation\cite{Deacon2017a} and skweness of the supercurrent profile.\cite{Sochnikov2015a}

In the last years, the quality of spin-orbit coupled quantum wires substantially increased.\cite{hg1,hg2} 
Moreover, a new generation of proximitized Rashba wires were fabricated that  exhibit a hard superconducting gap.\cite{Krogstrup2015a}
Some of these devices showed robust zero bias conductance peaks,\cite{Zhang2016a, Deng2017a}
and others allowed to explore excitation energy oscillations produced by an external magnetic field.\cite{Albrecht2016a}

In this article, we will focus on the study of conductance 
oscillations that arise in the Majorana-Rashba wire.
It is well established,\cite{Cheng2009a, Prada2012a,Rainis2013a, DasSarma2012a} that the origin of these oscillations resides in
the spatial overlap between the MBSs typically located at the ends of the wire: 
The MBS wave functions exhibit an oscillatory exponential decay towards the center. 
In the limit of high magnetic fields, the finite energy resulting from the overlap between 
the modes is approximately given by\cite{Cheng2009a, DasSarma2012a}
\begin{align}
\Delta E\approx \frac{\hbar^2k_\text{F,eff}}{m \chi}\cos\left(k_\text{F,eff} L\right) \exp\left(-\frac{2 L}{\chi}\right),
\label{eq:Eoverlap}
\end{align}
where $k_\text{F,eff}$, $L$, and $\chi$ are the effective Fermi wave vector, the length of the wire and the localization length of the MBS, respectively.
Due to the fact that $k_\text{F,eff}$ and $\chi$ increase with the 
magnetic field, the resulting overlap, and hence the conductance,  
should exhibit an oscillatory pattern with an increasing amplitude.

Recent experiments,\cite{Albrecht2016a} performed in Coulomb blockade Majorana islands\cite{Fu2010a, Hutzen2012a,Aguado2017a} show, however, clear deviations from this picture:
For an increasing magnetic field, most samples experience a decaying amplitude of the oscillations, resulting into crossings and anticrossings. 
On top of that, some samples feature that oscillations remain pinned at zero energy for a wide range of magnetic field ($\sim40\,$mT).
Furthermore, other samples manifest a vanishing conductance at high magnetic fields. 
\begin{figure}[tb]
\begin{center}
\includegraphics[width=1.\linewidth]{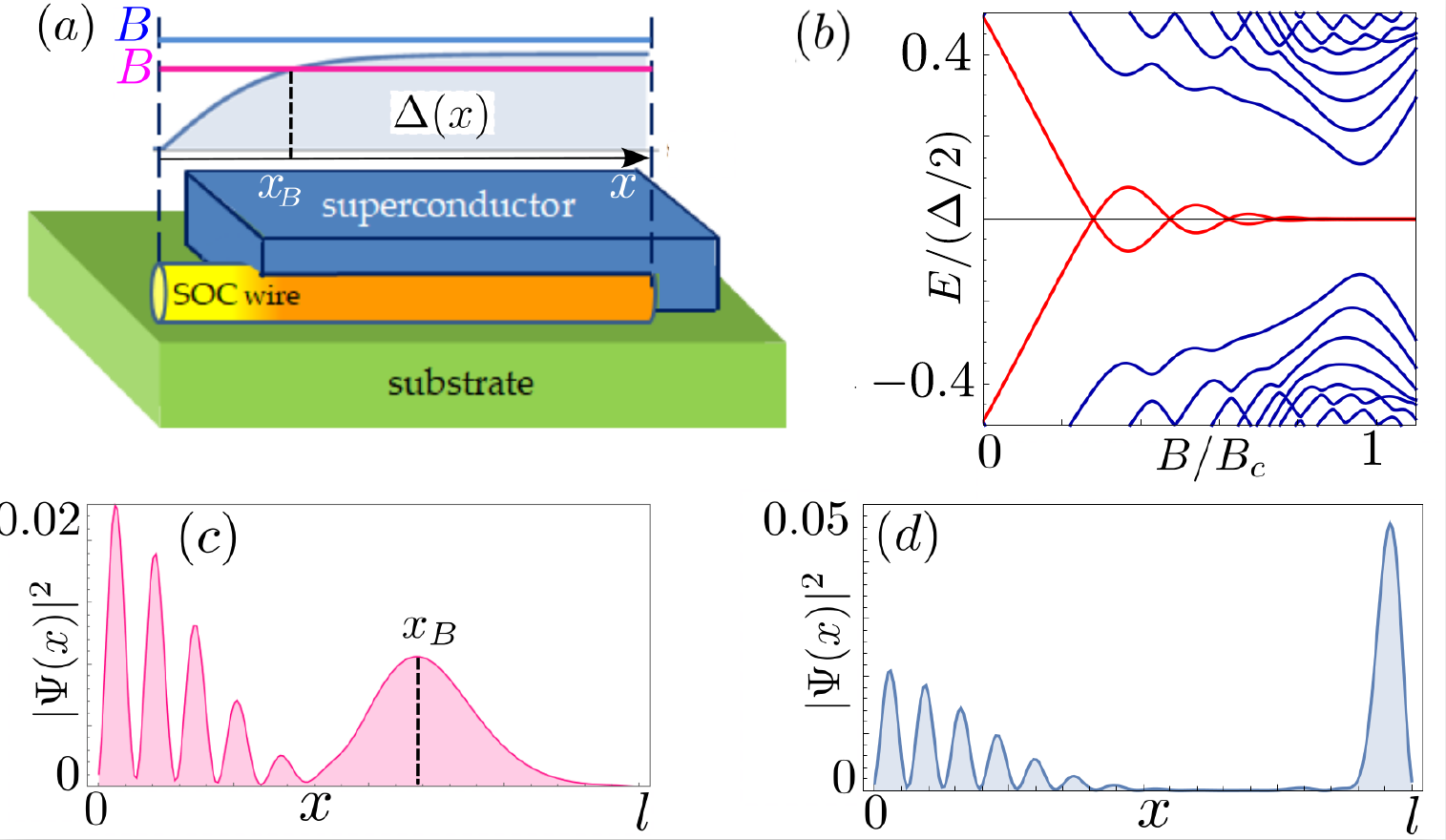}
	\caption{$(a)$ Schematic of the system. The spin-orbit coupled wire is placed on top of a substrate and partially covered by a superconductor. We assume a space dependent proximity induced pairing amplitude $\Delta(x)$. $(b)$ Numerical tight-binding calculation of the lowest energy eigenvalues, as a function of the Zeeman energy $B$ for $N=200$, $L=2~\mathrm{\mu m}$, $\alpha=-20.2$ meVnm, $\Delta_0=1.26$ meV, $\mu=0$, $\xi=0.8~\mathrm{\mu m}$. $(c)$ Density of the corresponding eigenfunctions for $B=0.9~B_c$, and $(d)$ $B=1.1 B_c$ as a function of $x$. In $(c)$ and $(d)$ the unspecified parameters have the same value as in $(b)$.}
	\label{Fig:profile}
\end{center}
\end{figure}
Motivated by these experimental results, some theoretical approaches 
introduced extra features into the original model:\cite{Lutchyn2010a,Oreg2010a} 
Adding Coulomb interactions between the electrons in the wire and the dielectric environment leads to zero energy pinning.\cite{Dominguez2017a} 
Including leakage current effects, coming from the presence of a drain in the superconductor, this leads to a vanishing conductance.\cite{Danon2017a}
Finally, the emergence of decaying oscillations can be obtained, taking into account
orbital effects,\cite{Klinovaja2017a} or wires with multiple occupied subbands,  high temperature, and 
simultaneous presence of Andreev bound states and MBSs.\cite{Chiu2017a, Liu2017a}

We, instead, study a simple scenario how topological decaying oscillations (see Fig.~\ref{Fig:profile}(b)) can appear: 
We introduce a finite coherence length in the superconducting pairing, 
that is,
\begin{align}
\Delta(x)=\Delta_0 \tanh(x/\xi),
\label{pairing}
\end{align}
where $\xi$ is the coherence length of the superconductor
\cite{Prada2012a, Stanescu2012a, Rainis2013a, Osca2013a} (see Fig.~\ref{Fig:profile}(a)).
Such a model is appropriate in a wide range of experimentally relevant situations. Indeed, in genuinely one-dimensional problems, such as the one we aim to describe, the superconducting pairing potential varies on length scales comparable to the coherence length as the geometrical end of the superconductor is approached.\cite{Volkov1974a,*Volkov1974b, Likharev1979a,Klapwijk2004a} The one-dimensional character of the physical setup is plausible, for instance, when superconductivity is induced by coating the nanowire with a thin film. Moreover, a smooth pairing potential is expected to be present if atoms of the coating are diffusing into the wire. In the latter case, however, the length $\xi$ is not directly related to the coherence length of the superconductor.\cite{Golubov2004a} The strength of the induced gap $\Delta_0$ does not only depend on the bare gap of the superconductor, but also on the contact between the wire and the Al film. To take all these options into account, we hence keep $\Delta_0$ and $\xi$ as independent variables.
Under such a pairing potential, the critical field for observing MBSs
reduces from $B_\text{c}=\sqrt{\Delta_0^2+\mu^2}$ to $B\approx \vert\mu\vert$ with $\mu$ the chemical potential.\cite{Prada2012a}
When the wire is globally in the topological phase, the Majorana fermions are located close to the left and right ends of the wire (see Fig.~\ref{Fig:profile}(d)). When, on the other hand, $\vert\mu\vert<B<B_{c}$, two MBS arise, placed close to the left end of the wire and the position $x_B$ satisfying $B=\sqrt{\Delta(x_B)^2+\mu^2}$ (see Fig.~\ref{Fig:profile}(c)).
It is interesting to note that in this case the magnetic field shifts the distance between MBSs, and thus, the maximum overlap between them decreases, which is reflected in their spectrum (see Fig.~\ref{Fig:profile}(b)).
Although some numerical results along these lines have been presented in Ref.~\onlinecite{Prada2012a}, new experimental results 
motivate a more careful analysis and understanding of a position dependent pairing. 
Here, we study numerically and analytically the shape of MBS wave functions 
arising below the critical bulk field $\vert\mu\vert \lesssim B <B_\text{c}$ for an arbitrary coherence length $\xi$. 
In striking contrast to the constant pairing scenario, 
we find two different Majorana fermion solutions with different $k$ space structure. 
A decaying oscillatory wave function placed close to the left end of the wire and a gaussian-like wave function placed at $x_B$ characterize the system.
The difference in the nature of the two Majorana fermions crucially influences their overlap, which, as $\alpha$, $\mu$ and $\xi$ are varied, can result in decaying oscillations, anticrossings or asymptotic decrease. All three scenarios have all been observed in experiments. As a further analysis of the properties of the model, we calculate the local linear conductance $G$ as a function of the applied magnetic field. We find that, in correspondence to the crossings and anticrossings in the lowest lying eigenvalues, $G$ develops non-quantized peaks. Interestingly, whenever $\xi$ is non-zero, the sharp transition between $G=0$ and $G=2e^2/h$, routinely associated to the topological phase transition, takes place for magnetic fields smaller than the value needed for the bulk topological phase transition. This behavior is in accordance with the above mentioned possibility of having Majorana bound states before the topological phase transition.

The outline of the paper is as follows: 
In Sec.~\ref{s2} we present the Majorana-Rashba model.\cite{Lutchyn2010a,Oreg2010a} 
Then, in Sec.~\ref{s3}, we discuss qualitatively the main results. In Sec.~\ref{sG}, we complement the qualitative analysis with quantitative calculations of the differential conductance. In Sec.~\ref{s4}, we present analytical approaches to the problem and carefully characterize the oscillations as a function of the Rashba spin-orbit coupling strength and the chemical potential. Finally, we conclude in Sec.~\ref{s5}

\section{Model}
\label{s2}
We study the Hamiltonian presented in Refs.~[\onlinecite{Lutchyn2010a}, \onlinecite{Oreg2010a}] 
${H}_c=\frac{1}{2}\int_0^{L} dx {\Psi}^{\dagger}(x)\mathcal{{H}}(x){\Psi}(x)$ with
\begin{eqnarray}
\mathcal{{H}}(x)&=&\left(\frac{-\partial_x^2}{2m^*}-\mu\right)\tau_z\otimes\sigma_0\nonumber\\
&-&i\alpha\partial_x\tau_z\otimes\sigma_z+B\tau_z\otimes\sigma_x+\Delta(x)\tau_x\otimes\sigma_z,
\label{Eq:Hamilton}
\end{eqnarray}
where $\Psi^{\dagger}(x)=[\psi_{\uparrow}^{\dagger}(x),\psi_{\downarrow}^{\dagger}(x),\psi_{\downarrow}(x),\psi_{\uparrow}(x)]$. The operators $\psi_{\uparrow,\downarrow}(x)$ annihilate a $\uparrow/\downarrow$ particle at position $x$ and the Pauli matrices $\sigma_i,~\tau_i$ with $i\in\{x,y,z\}$ act on spin- and particle-hole-space, respectively. 
In addition, $B=\frac{1}{2} g \mu_B B_x$ is the Zeeman energy, originating from a magnetic field applied in the $x$-direction $B_x$ ($B>0$ throughout the article), $m^*= 0.015 m_e$ if we choose the InSb effective mass, and $\mu$ is the chemical potential.  
The pairing potential $\Delta(x)$ is given by Eq.~\eqref{pairing}.

Using standard finite difference methods, we discretize Eq.~\eqref{Eq:Hamilton} yielding a $4N\times 4N$ matrix, henceforth called $\hat{H}_w$.
Here, we use $N$ for the total number of sites, 
and thus, $L=a_0 N$ is the length of the wire with $a_0$ the lattice spacing. 
This Hamiltonian has $4N$ eigenstates, denoted as $\psi^\nu(x)= (u_\uparrow^\nu,u_\downarrow^\nu,v_\downarrow^\nu,v_\uparrow^\nu)$ with corresponding eigenvalues $\epsilon_\nu$.
Conductance calculations are obtained by coupling the Majorana wire to normal contacts at each end. 
We account for the coupling to the leads by adding a constant and diagonal self-energy 
$\hat{\Sigma}_\text{L,R}^{r/a}=\pm i \hat{\Gamma}_\text{L,R}$, where
\begin{align}
&\hat{\Gamma}_\text{L}=\gamma_\text{L} \text{diag}(\hat{1},\hat{0},\hat{0},\cdots,\hat{0})_N,\label{gammaL}\\
&\hat{\Gamma}_\text{R}=\gamma_\text{R} \text{diag}(\hat{0},\hat{0},\hat{0},\cdots,\hat{1})_N.\label{gammaR}
\end{align}
Here, $\hat{1}$ and $\hat{0}$, denote the identity and zero $4\times 4$ matrices. 
The subindex $N$ refers to the number of $4\times 4$ matrix entries, yielding a
$4N\times 4N$ matrix $\hat{\Gamma}_\text{L,R}$.
Here, $\gamma_l=\pi \rho_l t_l^2$ is the broadening of the level coupled to the 
normal lead, and $\rho_l$ is the density of states of the $l$-lead, and $t_l$ a coupling constant.

Thus, we can construct the retarded and advanced Green's function of the open system as
\begin{align}
\mathcal{G}^{r/a}(\omega)= \lim_{\eta\to \pm 0}[\omega-\hat{H}_w- \hat{\Sigma}_L^{r/a} - \hat{\Sigma}_R^{r/a}+ i\eta]^{-1},
\end{align}
where $\mathcal{G}^{a}=(\mathcal{G}^{r})^\dagger$.
Using Keldysh techniques,\cite{Cuevas1996a} and assuming 
a negligible quasiparticle contribution, one can express the zero bias conductance in the $l$-lead as 
\begin{align}
\label{Eq:Cond}
G_l= \frac{2e^2}{h} (T_{\text{LAR},l}+T_{\text{CAR},l}),
\end{align} 
where 
\begin{align}
&T_{\text{LAR},l}= 4\text{Tr}\left[\hat{\Gamma}_{l}^e \mathcal{G}^r(0)\hat{\Gamma}_{l}^h \mathcal{G}^a(0)\right],\\
&T_{\text{CAR},l}=4\text{Tr}\left[\hat{\Gamma}_{l}^e \mathcal{G}^r(0)\hat{\Gamma}_{\overline{l}}^h \mathcal{G}^a(0)\right],
\end{align}
are the  local and crossed Andreev reflection at the $l$-lead, respectively.
Here, $l=\text{L,R}$ and $\overline{l}=\text{R,L}$. 
Besides, $\hat{\Gamma}_{l}^{e/h}$ are $4N\times 4N$ matrices, keeping only the electron/hole-like contributions of Eqs.~\eqref{gammaL} and~\eqref{gammaR}.

\section{Main results}
\label{s3}

\begin{figure}[tb]
	\begin{center}
	\includegraphics[width=1\linewidth]{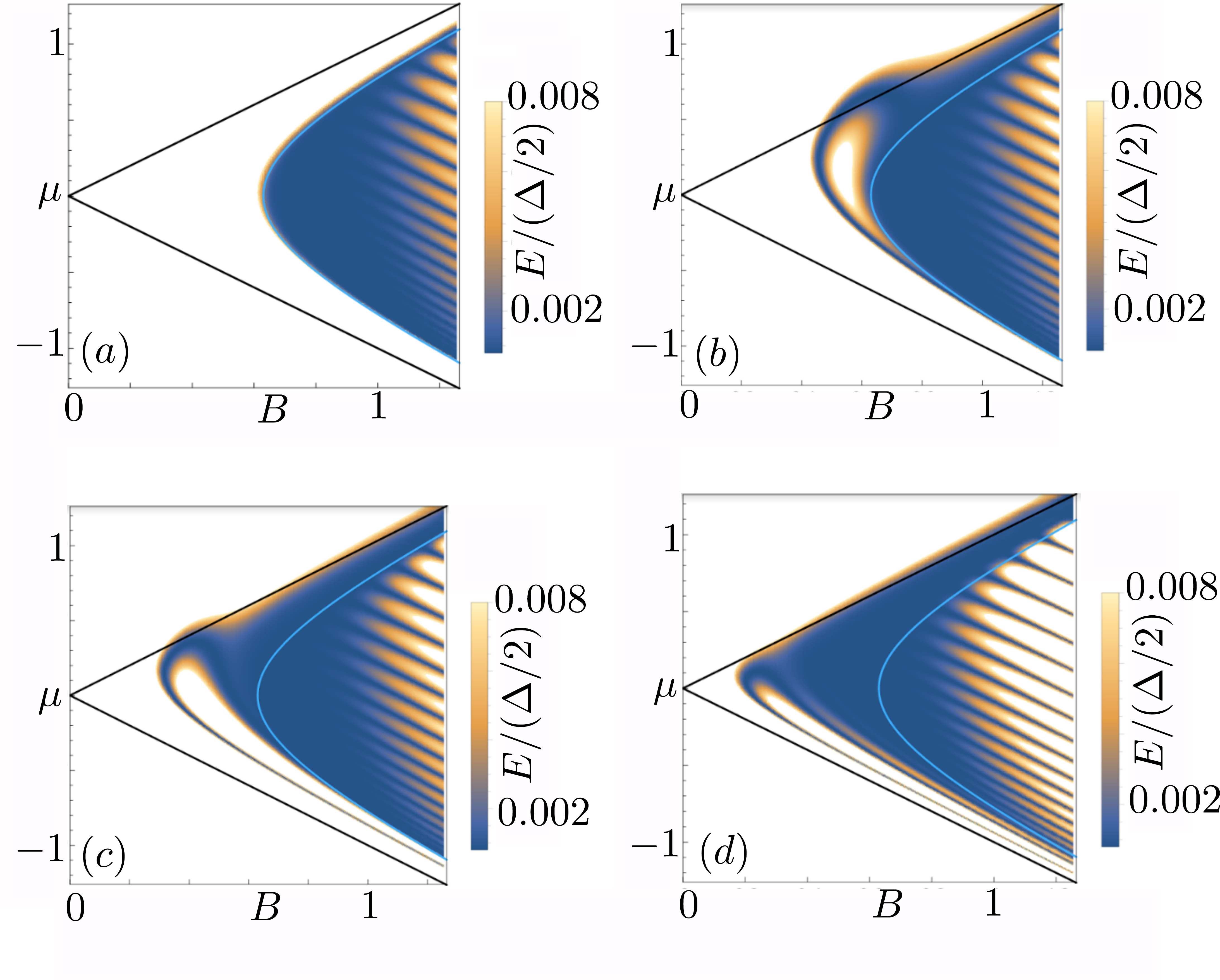}
		\caption{Numerical results for the lowest energy eigenvalues of the spin-orbit coupled wire as a function of Zeeman energy $B$ and chemical potential $\mu$ in meV. 
		The calculations are done for $N=200$, $L=2~\mathrm{\mu m}$, $\alpha=-20.2$ meVnm, $\Delta_0=0.63$ meV and different values for the coherence length $\xi$: (a) $\xi=10$ nm, (b) $\xi=200$ nm, (c) $\xi=400$ nm and (d) $\xi=1~\mathrm{\mu m}$. We highlight the lines $B=\vert\mu\vert$ and $\mu=\pm \sqrt{B^2-\Delta^2}$ in black and blue, respectively.}
		\label{fig:e_B_mu}
	\end{center}
\end{figure}

In this section, we characterize the effects that a finite coherence length 
introduce in the critical field and the oscillating pattern resulting from 
the hybridization of MBSs.
To this aim, we diagonalize the discretized version of Eq.~\eqref{Eq:Hamilton} and
compare in Fig.~\ref{fig:e_B_mu}
the lowest energy states in the parameter space $(B,\mu)$ 
for different coherence lengths: $\xi=10$nm, $\xi=200$nm, $\xi=400$nm and $\xi=1\mu$m.
We can observe that for $B>\sqrt{\Delta_0^2+\mu^2}\equiv B_{c}$ (see blue curve in Fig.~\ref{fig:e_B_mu})
the qualitative topological properties of the Rashba-wire
are still present, i.e.~MBSs localized close to the two ends of the wire arise and 
oscillate with increasing amplitude for increasing magnetic fields. 

For $B<B_{c}$, MBSs arise for an increasing $\xi$, see Fig.~\ref{fig:e_B_mu}(a)-(d).
The reason for this appearance can be understood if we consider a slowly varying pairing potential. 
In this situation, the critical condition $B_\text{c}(x)=\sqrt{\Delta(x)^2+\mu^2}$
can be satisfied locally, and thus, 
two MBSs arise: One is placed close to the left end of the wire, $x_\nu\sim 0$, and another one at $x_B$, where the relation $B=\sqrt{\Delta(x_B)^2+\mu^2}$ is satisfied. 
Note that $x_B$, and thus, the distance between the MBSs, increases for an increasing magnetic field.
Roughly speaking, the requirement for having Majorana fermions 
is hence no longer $B>B_{\text{c}}$, but becomes related to the existence of the point $x_B$, 
that is guaranteed to exist for $B>B_{\mu}\equiv |\mu|$.
This behavior is indeed what we observe in Fig.~\ref{fig:e_B_mu}: For an increasing coherence length, 
zero energy states approach asymptotically to $B=\vert\mu\vert$ (see black curves in Fig.~\ref{fig:e_B_mu}).
Interestingly, this means that MBSs are present whenever the system is 
in the quasi-helical regime of the spin-orbit coupled wire,\cite{Streda2003a,Meng2013a,Gambetta2015a,Gambetta2014a} that is, 
whenever the system in the absence of superconductivity is effectively spinless. 
We observe deviations from this behavior for shorter coherence lengths (see Fig.~\ref{fig:e_B_mu} (b)), and zero-energy states can arise even for $B<\vert\mu\vert$. 

\begin{figure}
\includegraphics[scale=0.13]{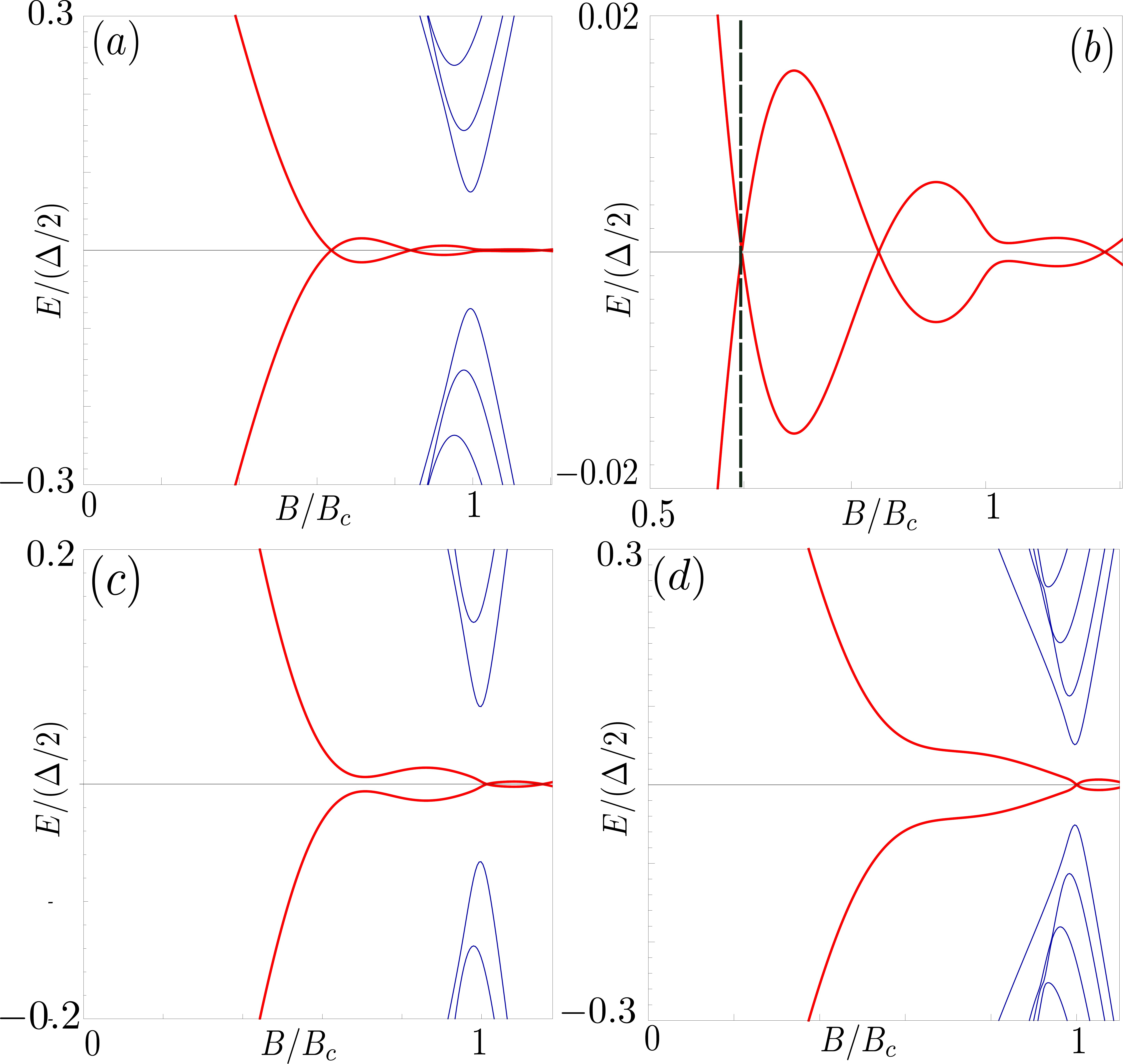}
	\caption{Numerical results of the low energy eigenvalues of a spin-orbit coupled wire of total length $N=200$, $L=2~\mathrm{\mu m}$ with $\alpha=-22.7$ meVnm, $\Delta_0=0.76$ meV, $\xi=0.2~\mathrm{\mu m}$ and different chemical potential: (a)-(b) $\mu=0.63$ meV, (c) $\mu=0.76$ meV, (d) $\mu=1.01$ meV. (a) and (b) have the same parameters with different scaling of the axes. The vertical dashed line in (b) represents $B=\vert\mu\vert$.}
	\label{Fig:MainResult1}
\end{figure}

\begin{figure}[tb]
\begin{center}                                    
\includegraphics[width=1\linewidth]{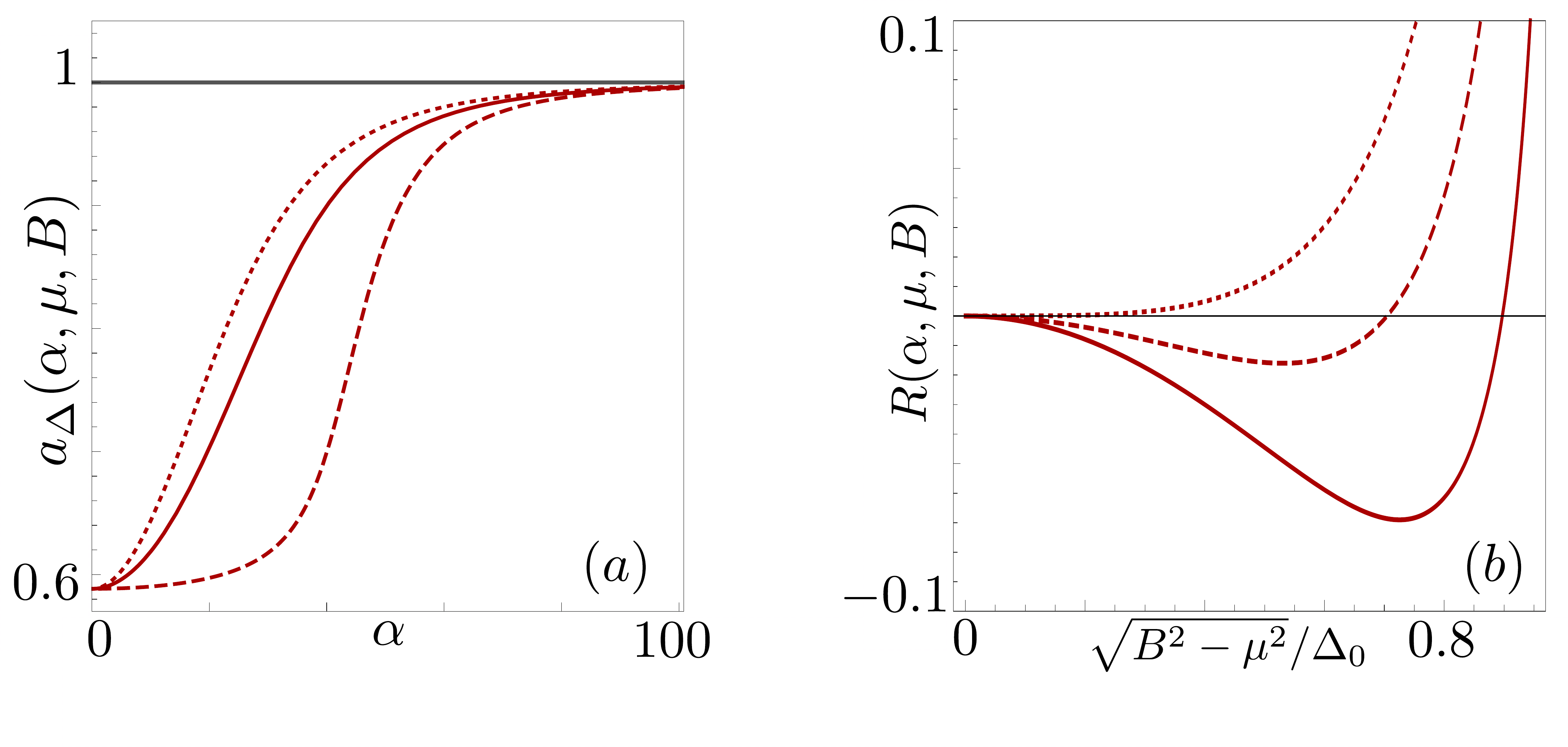}
	\caption{$(a)$ $a_{\Delta}(\alpha,\mu,B)$ as a function of spin-orbit coupling $\alpha$ in meVnm with $B=0.35$ meV, $\Delta_0=0.63$ meV and $\mu=0.33$ meV (dotted), $\mu=-0.33$ meV (dashed), $\mu=0$ (solid). $(b)$ $R(\alpha,\mu,B)$ for 3 different values of $a$: $a_{\Delta}(\alpha,\mu,B)=1$ (dotted), $a_{\Delta}(\alpha,\mu,B)=0.8$ (dashed), $a_{\Delta}(\alpha,\mu,B)=0.6$ (solid).}
	\label{Fig:Relation}
\end{center}
\end{figure}

The existence of MBSs below $B_{\text{c}}$ is not the only interesting effect of a finite coherence length. 
The dependence that the lowest energy level has as a function of the applied magnetic field is also remarkable. 
Since the distance between the two Majorana fermions increases when the magnetic field is increased, 
the resulting overlap decreases, 
see Figs.~\ref{Fig:profile} (b) and~\ref{Fig:MainResult1} (a)-(b). 
This feature is often observed in experiments and is difficult to interpret. 
However, in the context of a finite coherence length, decaying oscillations for $B<B_{\text{c}}$ appear naturally.
In this scenario, decaying oscillations are, interestingly, just one of the possible behaviors. 
It is worth to notice 
that decaying oscillations (Fig.~\ref{Fig:MainResult1} (a)) can evolve into anticrossings 
(Fig.~\ref{Fig:MainResult1} (c)) and finally into an monotonic decay to zero (Fig.~\ref{Fig:MainResult1} (d)) as the chemical potential or the 
spin-orbit coupling are increased.
In order to understand the microscopic mechanisms that induce the different patterns, we analyze
the physical properties of the lowest energy BdG wavefunctions. The Majorana fermion around $x_B$ is expected to be a non-oscillating function of $x$, in particular for $\mu=0$, a Gaussian \cite{Oreg2010a}. An oscillating hybridization energy can hence only emerge from an oscillating wavefunction of the Majorana fermion located at $x_{\nu}$.
For a qualitative discussion of the wavefunction around $x_{\nu}$, we start by considering the case of a constant superconducting pairing. 
In this picture, MBSs mainly have dominant contributions from momenta 
$k$ around $k\equiv k_0=0$ and $k=\pm k_F$. Thus, the MBS wave function can be expressed as 
the linear combination $\psi\sim \psi_{k_0}+\psi_{k_\text{F}}+\psi_{-k_\text{F}}$.\cite{Klinovaja2012a} 
All these contributions have a spinor structure and decay exponentially, 
with a typical localization length related to the corresponding direct energy gap, that is,
\begin{align}
\psi_{j} \propto \exp(-\Delta_{j} x+ i k_j x),
\end{align}
where the index $j$ resembles $k_0$ and $~\pm k_\text{F}$.
Around $k_0$ the gap is given by $\Delta_{k_0}=\sqrt{B^2-\mu^2}-\Delta_0$, and does not depend on $\alpha$. 
In contrast, at $k=\pm k_F$, the gap is given by a fraction of the bare induced superconducting 
coupling $\Delta_{\pm k_F}=a_\Delta(\alpha,\mu,B)\Delta_0$. 
The behavior of $a_\Delta(\alpha,\mu,B)$ as a function of $\alpha$, for different values of $\mu$ is depicted in Fig.~\ref{Fig:Relation} (a)\cite{San-Jose2012a}.
We can observe that as $\alpha$ or $\mu$ are decreased, $a(\alpha,\mu,B)$ and hence $\Delta_{\pm k_F}$ decrease. 
Note in passing that the contributions $\psi_{\pm k_\text{F}}$, oscillate with the wave vector $k_\text{F}$.
Therefore, whenever $\Delta_{\pm k_F}/\Delta_{k_0}< 1$, the wave function will exhibit a spatial oscillatory pattern.

In the limit of a slowly varying pairing potential, i. e. $k_F\gg 1/\xi$, it is possible to find similar expressions as in the constant pairing case (see Sec.~\ref{s4}). Due to the position dependent pairing, the exponents are then replaced by
\begin{align}
 \Delta_{j} x\rightarrow \int_0^{x} \Delta_{j}(x') dx',
\end{align}
where $\Delta_{j}(x)$ is the result of substituting $\Delta(x)$ into the direct gap expressions.
An oscillatory pattern of the hybridization energy is observed when the oscillating contribution of the wavefunction located at $x_{\nu}$ is dominant around the location of the second Majorana ($x=x_B$). This condition is indeed fulfilled whenever 
\begin{equation}
\frac{
\int_0^{x_B} \Delta_{k_\text{F}}(x) dx}{\int_0^{x_B} \Delta_{k_0}(x) dx}< 1.
\end{equation}
As shown in Fig.~\ref{Fig:Relation} (b) (see also Eq. \eqref{Eq:Relation5} for more details), the relation is satisfied for small values of $a_{\Delta}(\alpha,\mu,B)$, 
i.e.~weak spin-orbit coupling and/or small $\mu$ close to the topological phase transition. 
For strong spin-orbit coupling and large $\mu$, however, the relation does not hold anymore and the oscillations disappear.
A deeper analysis of the wavefunctions, going beyond simple scaling arguments, is presented in Sec.~\ref{s4}.
Note that some physical behaviors described in this section could be 
transposed into the scenario where the superconducting pairing is roughly constant while the chemical potential acquires a spatial dependence due to, for instance, the presence of smooth confinement.\cite{Prada2012a, Kells2012a, CMoore2018a}

\section{Conductance}
\label{sG}

We complement the analysis of the previous section with a quantitative calculation of the zero bias conductance $G_\text{L}:=G$ in the notation of Eq. (\ref{Eq:Cond}).
In the following, we will analyse two different scenarios: crossings/anticrossings and the asymptotic approach to zero.

\begin{figure}[tb]
	\begin{center}
\includegraphics[width=1\linewidth]{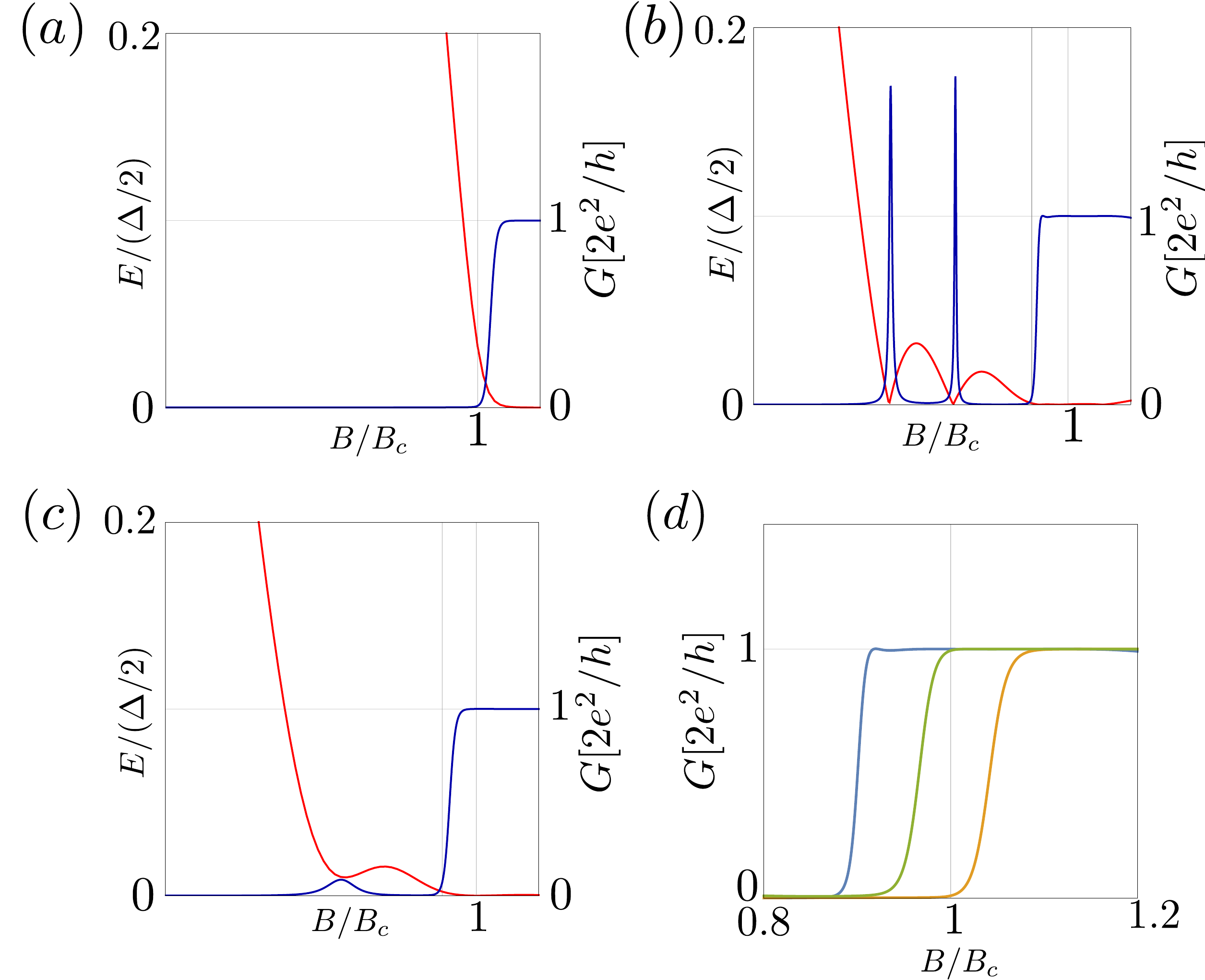}
\caption{(a)-(c) Lowest energy eigenvalue of the system (red, left vertical axis) and zero-energy conductance (blue, right vertical axis) as a function of  $B/B_c$ with the parameters $N=180$, $L=1.8~\mathrm{\mu m}$, $\Delta_0=0.66$ meV, $\mu=0.2$ meV. Further $\xi=0$ in (a), while $\xi=0.5~\mathrm{\mu m}$ for (b)-(c), as well $\alpha=20.16~$meVnm in (a), (b) and (d), while $\alpha=25.2~$meVnm in (c). In (d), we illustrate the conductance calculation of (a) (yellow), (b) (blue) and a third one with the parameters of (c) but $\xi=0.25~\mathrm{\mu m}$. Furthermore, $\gamma_L=\gamma_R=2.52~$meV.}
\label{Fig:cond}
\end{center}
\end{figure}

{\it Crossings/anticrossings}. ---
Anticrossings with an energy gap $\delta\epsilon$, give rise to $hG/(2e^2)\sim 2\gamma_L^2/(\delta\epsilon)^2\ll 1$, for $\delta \epsilon\gg \gamma_\text{L}$ (see Fig.~\ref{Fig:cond}). 
In turn, the crossing points exhibit conductance peaks with values between $4e^2/h<G<2e^2/h$, see Fig.~\ref{Fig:cond}(b). Evidently, each MBS can  contribute to the conductance since both MBS wave functions exhibit a finite weight at $x=0$\cite{Prada2017a}. 
In this situation, the left (right) MBS is (not) perfectly spin polarized.\cite{Sticlet2012a,Prada2017a} This is known as a reason why it contributes to the conductance with $G\approx 2e^2/h$ ($G\lesssim 2e^2/h$),
yielding a total subgap conductance of $G<4e^2/h$. This situation is similar to the crossings that can emerge when a 
non-proximitized part is added to the system before the leads.\cite{Cayao2015a}

{\it Asymptotic decay to zero}. ---
The magnetic field shifts the MBS placed at $x_\text{B}$, yielding an exponential reduction of the energy. 
When the energy spectrum reaches zero, the conductance jumps abruptly to the quantized value $G=2e^2/h$, even for $B<B_\text{c}$.
Interestingly, as $\xi$ is increased, the transition shifts towards smaller values of $B$, see Fig.~\ref{Fig:cond}$(d)$.
The behavior discussed in this section is consistent with the qualitative discussion given in Sec.~\ref{s3} and with the analytical results given in the following section.

\section{Analytical analysis}
\label{s4}

In order to derive simple relations explaining the behavior observed by means of the numerical solution, we are inspired by Ref. \onlinecite{Klinovaja2012a} and simplify the model in two regimes: strong and weak spin orbit coupling. Beyond Ref. \onlinecite{Klinovaja2012a}, we will then solve the models in the presence of a non-uniform pairing potential.

\subsection{Effective model for strong spin-orbit coupling}
\begin{figure}
\includegraphics[scale=0.5]{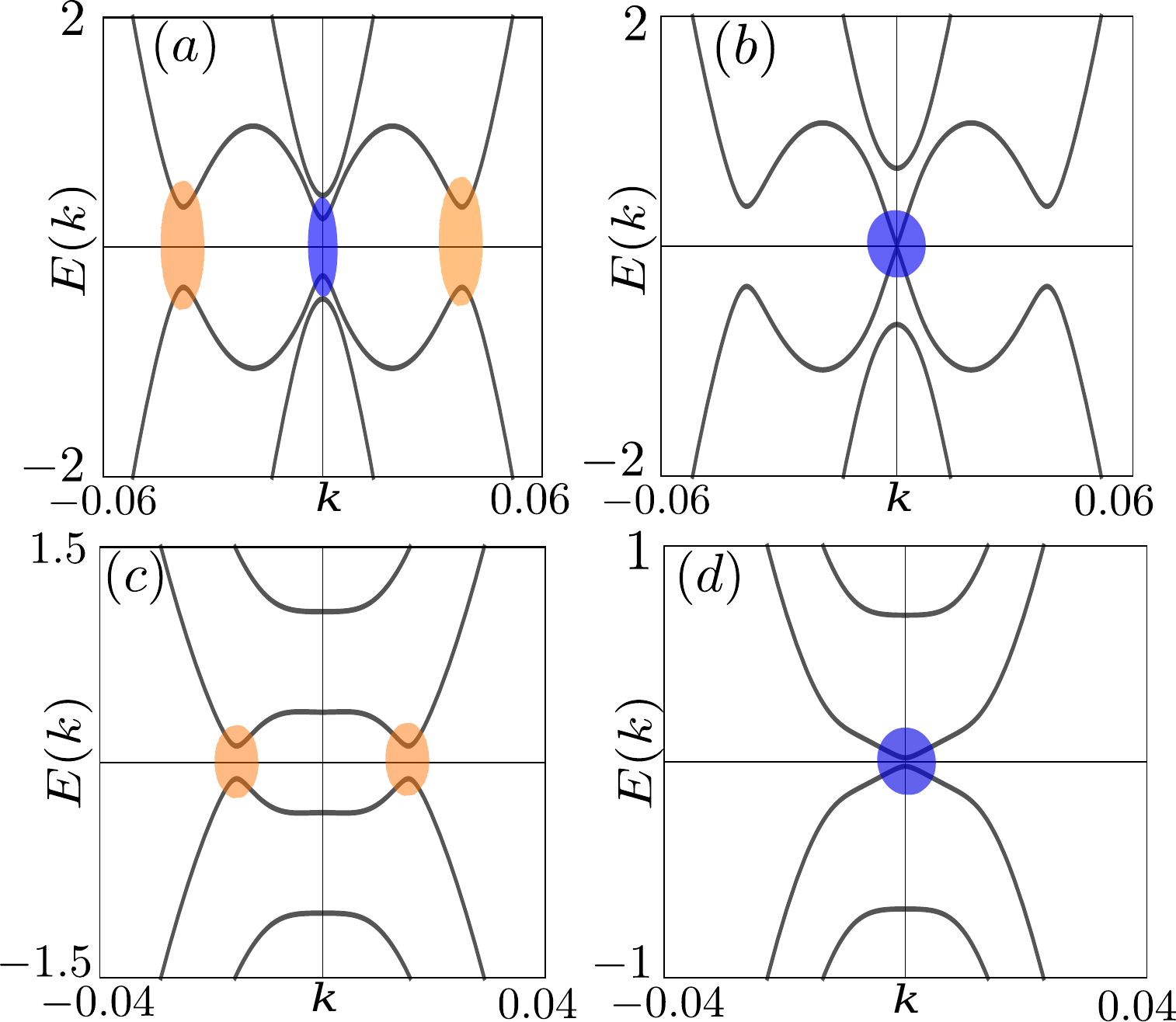}
\caption{Dispersion relation of the full continuum model with $E(k)$ in meV and $k$ in $\mathrm{nm}^{-1}$ in the strong spin-orbit regime ((a)-(b)) with $\alpha=-100$ meVnm, $\Delta_0=0.35$ meV, $\mu=0$ and (a) $B=0.1$ meV, (b) $B=0.33$ meV and weak spin-orbit regime ((c)-(d)) with $\alpha=-15$ meVnm, $\Delta_0=0.35$ meV, $\mu=0$ and (c) $B=0.7$ meV, (d) $B=0.33$ meV.}
\label{Fig:dispersion}
\end{figure}

In the strong spin-orbit coupling regime, $m^*\alpha^2 \gg B,~ \Delta_0$, we complement the continuum model with the assumptions of a slowly varying superconducting pairing potential $m^*\alpha \gg 1/\xi$. Then, the continuum Hamiltonian can be further simplified to effective (linear) Hamiltonians around zero average momentum (i) and momentum $k_F\cong 2m^*\alpha$ (ii) (see also Fig. \ref{Fig:dispersion} (a)-(b) for a schematic):
\begin{equation}
\mathrm{(i)}~~\langle -i\partial_x\rangle\simeq 0, ~~~~~ \mathrm{(ii)} ~~\langle -i\partial_x\rangle\simeq \pm  k_F.
\end{equation}
For case (i), we are allowed to neglect the quadratic part of Eq.~(\ref{Eq:Hamilton}), resulting in the low energy Hamiltonian
\begin{eqnarray}
{H}_I&=&\int_0^{l} dx {\Psi}^{\dagger}(x)\bigg[-i\alpha\partial_x\tau_z\otimes\sigma_z\nonumber\\
&-&\mu\tau_z\otimes\sigma_0+B\tau_z\otimes\sigma_x+\Delta(x)\tau_x\otimes\sigma_z\bigg]{\Psi}(x).
\label{Eq:ham3.1}
\end{eqnarray}
In case (ii), we can perform a spin-dependent gauge transformation
\begin{equation}
{\Psi}(x)=e^{-2im^*\alpha x(\tau_0\otimes\sigma_z)}\tilde{\Psi}(x),
\label{Eq:trafo}
\end{equation}
where $\tilde{\Psi}(x)$ is a slowly varying function of $x$ with respect to $1/(m*\alpha)$ at low energy. After plugging Eq.~(\ref{Eq:trafo}) into Eq.~(\ref{Eq:Hamilton}) and linearizing, the transformed Hamiltonian becomes
\begin{eqnarray}
{H}_E&=&\int_0^{l} dx \tilde{\Psi}^{\dagger}(x)\bigg[i\alpha\partial_x\tau_z\otimes\sigma_z\nonumber\\
&-&\mu\tau_z\otimes\sigma_0+a_{\Delta}(\alpha,\mu,B)\Delta(x)\tau_x\otimes\sigma_z\bigg]\tilde{\Psi}(x),
\label{Eq:ham3}
\end{eqnarray}
where the fast oscillating terms are integrated out.\\
Here, $a_{\Delta}(\alpha,\mu,B)\in \{0,1\}$ is determined by the dispersion relation of the lowest energy eigenvalue of Eq.~(\ref{Eq:Hamilton}) (assuming constant superconducting pairing $\Delta_0$). In case of strong spin orbit coupling, we obtain $a_{\Delta}(\alpha,\mu,B)\rightarrow 1$. The full behavior of $a_{\Delta}(\alpha,\mu,B)$ as a function of $\alpha$ is illustrated in Fig. (\ref{Fig:Relation}) (a). For small $\alpha$, we find the analytical expression
\begin{equation}
a_{\Delta}(\alpha,\mu,B)=\frac{\sqrt{\Delta_0^2+B^2}-B}{\Delta_0}+\frac{4m^{*2}(B+\mu)}{\Delta_0\sqrt{B^2+\Delta_0^2}}\alpha^2+\mathcal{O}(\alpha^3).
\label{Eq:aDelta}
\end{equation} 
The rational behind the approximation scheme leading to Eqs.~(\ref{Eq:ham3.1}) and (\ref{Eq:ham3}) is that, for strong spin-orbit coupling and for weak translational symmetry breaking by the applied superconducting pairing, we expect that the main effect of superconductivity is to renormalize the helical gap close to zero momentum and open a gap close to $k_F$, as schematically shown in Fig \ref{Fig:dispersion}. Any low-energy eigenstate is then evaluated as linear combination of eigenstates of ${H}_I$ and ${H}_E$.\\

\subsection{Effective model for weak spin-orbit coupling}
In case of weak spin-orbit coupling, $m^*\alpha^2\ll B$, we additionally assume $\sqrt{m^*B}\gg 1/\xi$. To develop effective linear models for our purposes, in this case, we explicitly distinguish two regimes in parameter space: deep inside the topological phase and close to the phase transition. Far away from any boundary, the latter case is hence described within the linear Hamiltonian of Eq.~(\ref{Eq:ham3.1}) (see Fig. \ref{Fig:dispersion} (d)), while, close to the boundaries, the contribution around $k=\pm k_F$ is still important. Deep inside the topological phase, the gap opened at $k=0$ is large compared to the gap opened at $k=\pm k_F$ given by $a_{\Delta}(\alpha,\mu,B)\Delta_0$ (see Fig. \ref{Fig:dispersion} (c)). For weak spin-orbit coupling, $a_{\Delta}(\alpha,\mu,B)\ll 1$, the low energy physics is described around the points $k=\pm k_F$. An appropriate linear model has to take into account that spins are not (quasi-)helical in the weak spin-orbit regime but acquire a spin-tilting. To implement this feature in the linear model we, demand an artificial $\epsilon$, $\gamma$ and $\nu$, acting as magnetic field, chemical potential and Fermi velocity. The momentum-space Hamiltonian in the absence of superconductivity is given in the spin-resolved basis as
\begin{equation}
\label{Eq:Hlin}
\mathcal{H}^0_{\text{lin}}(k)=\begin{pmatrix}
\nu k-\gamma & \epsilon\\
\epsilon & -\nu k-\gamma
\end{pmatrix}.
\end{equation}
Unlike former linear models of this paper, here we require $\gamma\geq \epsilon$ to make the model appropriate for our purpose. To connect those parameters to the physical parameters of the spin-orbit coupled wire, we demand three conditions to hold: (i) Fermi-surface, (ii) velocity at the Fermi points, and (iii) spin-tilting at the Fermi surface have to coincide in both models. The demands (i)-(iii) result in the following conditions
\begin{eqnarray}
k_{\text{F,lin}}&=&k_{\text{F,SOC}}\equiv k_{\text{F}},\nonumber\\
v_{\text{F,lin}}&=&v_{\text{F,SOC}}\equiv v_{\text{F}},\nonumber\\
\frac{\nu k_F+\sqrt{\nu ^2k_F^2+\epsilon^2}}{\epsilon}&=&\frac{k_F\alpha - \sqrt{B^2+ k_F^2\alpha^2}}{B},
\label{Eq:EquationSys}
\end{eqnarray}
where the last equation originates from the eigenvectors of Eq.~(\ref{Eq:Hamilton}) with $\Delta(x)=0$ and Eq.~(\ref{Eq:Hlin}). The equation system (\ref{Eq:EquationSys}) has the unique solution
\begin{eqnarray}
\label{Eq:Vel}
\nu &=&\frac{(1+\kappa^2)v_{F}}{-1+\kappa^2},~~\epsilon=\frac{2\kappa(1+\kappa^2)v_{F}k_{F}}{(-1+\kappa^2)^2},\nonumber\\~~\gamma &=&\frac{(1+\kappa^2)^2v_Fk_F}{(-1+\kappa^2)^2}
\end{eqnarray}
with the replacements
\begin{eqnarray}
\kappa &=&\frac{k_{F}\alpha-\sqrt{B^2+k_{F}^2\alpha^2}}{B},\nonumber\\
v_{F}&=& 4m^* k_{F}-\frac{k_{F}\alpha^2}{\sqrt{B^2+k_{F}^2\alpha^2}},\nonumber\\
k_{F}&=&\sqrt{2}\sqrt{m^{*2}\alpha^2+m^{*}\mu+m^{*}\sqrt{B^2+m^{*2}\alpha^4+2m^{*}\alpha^2\mu}}.\nonumber\\
\end{eqnarray}
A feature that does not coincides in both models is the spin rotation length along the dispersion relation. However, this feature only plays a minor role for spectroscopic properties. The direction of the spin rotation along the dispersion, however, is the same in both models. \\
Including superconducting pairing, the linearized model then becomes
\begin{eqnarray}
{H}_{\mathrm{lin}}&=&\int_0^{l} dx {\Psi}^{\dagger}(x)\bigg[i\alpha\partial_x\tau_z\otimes\sigma_z
-\gamma\tau_z\otimes\sigma_0\nonumber\\
&+&\epsilon\tau_z\otimes\sigma_x+a_{\Delta}(\alpha,\mu,B)\Delta(x)\tau_x\otimes\sigma_z\bigg]{\Psi}(x).
\label{Eq:ham3.2}
\end{eqnarray}
\subsection{Wave function at $x_B$ and new critical field}
\label{Sec:subC}
The existence of zero energy MBS in the trivial phase can be fully understood within this analytical approach. We first focus on the large $\xi$ limit. In this regime, we can apply the linear approximations of Eqs.~(\ref{Eq:ham3.1}), (\ref{Eq:ham3}) and (\ref{Eq:ham3.2}). As a starting point, we concentrate on a system with no boundaries and demand the existence of one point in space, $x=x_B$, which satisfies the relation $\Delta^2(x_B)=B^2-\mu^2$. Around this point, the low-energy physics (for strong and weak spin-orbit coupling) is described within the linear approximation of Eq.~(\ref{Eq:ham3.1}) only, since it becomes gapless. We, therefore, search for zero-energy solutions $\mathcal{{H}}_I(x)\Phi(x)=0$, where we demand $\Phi(x)$ to be of the form $\Phi(x)=U\chi(x)$, with
\begin{equation}
U=\frac{1}{\sqrt{2}}\left(\tau_0\otimes\sigma_0-i\tau_x\otimes\sigma_y\right).
\end{equation}
This unitary transformation reorganizes the Hamiltonian in the Majorana basis.
For $\chi(x)$ we further assume a solution of the form
\begin{equation}
\chi(x)=(a,b,c,d)^T\exp[f(x)].
\label{Eq:sol1}
\end{equation}
After plugging in this ansatz \cite{Timm2012a,Fleckenstein2016a,Traverso2017a,Porta2018a}, we obtain the solution
\begin{equation}
\label{Eq:f(x)}
f(x)=\pm \int dx \frac{\Delta(x)\pm \sqrt{B^2-\mu^2}}{\alpha}.
\end{equation}
In the case of linear behavior of $\Delta(x)$ with respect to $x$ and $\mu=0$, we restore the groundstate solution of the displaced harmonic oscillator, solved in Ref. \onlinecite{Oreg2010a}. With the solutions of Eq.~(\ref{Eq:f(x)}), we obtain the corresponding spinor structure of $\Phi(x)$
\begin{eqnarray}
(u,v,\tilde{v},\tilde{u})^T=U(a,b,c,d)^T=\nonumber\\\frac{1}{\sqrt{2}}\left( \pm e^{\pm i\varphi}+i,ie^{\pm i \varphi}\pm 1,-ie^{\pm i \varphi} \pm 1,\pm e^{\pm i \varphi}-i\right)^T,
\label{Eq:spinor}
\end{eqnarray}
where $\varphi= \arccos(B/\mu)$. To fulfil the Majorana condition $\tilde{u}=u^*$ and $\tilde{v}=v^*$, we need $\exp[i\varphi] \in \mathbb{R}$, which is only true for
\begin{equation}
B\geq |\mu|.
\label{Eq:Bmu}
\end{equation}
For a slowly varying $\Delta(x)$, the latter relation represents a bound for the formation of Majorana zero-modes. The hand-waving argument given in the previous section can hence be put on a formal basis.\\
Compiling Eqs.~(\ref{Eq:sol1}),~(\ref{Eq:f(x)}) and~(\ref{Eq:spinor}) and imposing that the wavefunction is normalizable and centered around $x_B$, we explicitly obtain
\begin{equation}
\Phi(x)=\frac{1}{\sqrt{N}}\begin{pmatrix}
e^{i\varphi}+i\\
ie^{i\varphi}+1\\
-ie^{i\varphi}+1\\
e^{i\varphi}-i
\end{pmatrix}e^{-\int_0^x dx' \frac{1}{\alpha}\left(\Delta(x')-\sqrt{B^2-\mu^2}\right)},
\label{Eq:Chi1}
\end{equation}
with the normalization constant $N$.
\subsection{Wavefunction at $x_\nu$ in the strong spin-orbit coupling regime}
\label{Sec:subD}
At the left end of the wire, where the proximity induced pairing decreases to zero, we have to distinguish between effective models of strong and weak spin-orbit coupling. For strong spin-orbit coupling, the low-energy physics is captured by a linear combination of eigenstates of Eqs.~(\ref{Eq:ham3.1})~and~(\ref{Eq:ham3})
\begin{equation}
\Psi(x)\simeq a_1\Psi_I(x)+b_1e^{-2i m^*\alpha x (\tau_0\otimes\sigma_z)}\Psi_E(x)
\end{equation}
with the coefficients $a_1$ and $b_1$ to be derived by the boundary conditions, and $\Psi_l(x)$, $l\in I,E$, satisfying
\begin{equation}
\mathcal{H}_l(x)\Psi_l(x)=0.
\label{Eq:sol2}
\end{equation}
The solution for $\Psi_I(x)$ is constructed by means of Eqs.~(\ref{Eq:sol1}),~(\ref{Eq:f(x)})~and~(\ref{Eq:spinor}). The solution for $\mathcal{H}_E(x)$ can be found in an analogous way after multiplying Eq.~(\ref{Eq:sol2}) (with $l=E$) from the left with $\tau_z\otimes\sigma_z$. Using the properties of Pauli matrices, especially $[\tau_y\otimes\sigma_0,\tau_0\otimes\sigma_z]=0$, where $[.,.]$ denotes the commutator, integration yields the solution
\begin{eqnarray}
\Psi_E(x)&=&\exp\bigg[\int_0^x dx'\frac{i}{\alpha}\bigg(\Delta(x')(\tau_y\otimes\sigma_0)\nonumber\\
&-&\mu(\tau_0\otimes\sigma_z)\bigg)\bigg]\Psi_0.
\label{Eq:PsiE}
\end{eqnarray}
Subsequently, the spinor $\Psi_0$ has to be chosen such that the wavefunction satisfies the Majorana condition $\Psi(x)=[u(x),v(x),v^*(x),u^*(x)]^T$, which results in four possible solutions. Furthermore, assuming a semi-infinite system ($x>0$), we have to satisfy the boundary condition $\Psi(0)=0$. Moreover, the solution has to decay away from $x=0$.
The first condition implies that the spinors $\Psi_I(0)$ and $\Psi_E(0)$ are linearly dependent. Hence, from Eq.~(\ref{Eq:PsiE}), we select the solutions for $\Psi_E(x)$ which decay away from $x=0$ and combine them with their linearly dependent counterparts $\Psi_I(x)$. This leads to the only physical solution \begin{eqnarray}
\Psi(x)&=&\frac{1}{\sqrt{N}}
\begin{pmatrix}
i-e^{i\varphi}\\
ie^{i\varphi}-1\\
-ie^{i\varphi}-1\\
-e^{i\varphi}-i
\end{pmatrix}e^{\int_0^x dx'\frac{1}{\alpha}\left(\Delta(x')-\sqrt{B^2-\mu^2}\right)}\nonumber\\
&-&\frac{1}{\sqrt{N}}e^{-i(2m^*\alpha+\frac{\mu}{\alpha})x(\tau_0\otimes\sigma_z)}
\begin{pmatrix}
i-e^{i\varphi}\\
ie^{i\varphi}-1\\
-ie^{i\varphi}-1\\
-e^{i\varphi}-i
\end{pmatrix}
e^{-\int_0^x dx'\frac{\Delta(x')}{\alpha}}\nonumber\\
\label{Eq:MajoranaSol}
\end{eqnarray}
with normalization constant $N$.
\subsection{Wavefunction at $x_\nu$ in the weak spin-orbit coupling regime}
\label{Sec:subE}
The wave function at $x_\nu$ for the case of weak spin-orbit coupling  ($a_{\Delta}(\alpha,\mu,B)\ll 1$) is given by linear combination of eigenstates of Eq.~(\ref{Eq:ham3.2}), which indeed have the same form as the eigenstates of Eq.~(\ref{Eq:ham3.1}) with the replacements $B\rightarrow\epsilon$, $\mu\rightarrow\gamma$ and $\alpha\rightarrow-\nu$. A consequence of neglecting all other contributions in this linear approach is that we can only accomplish the boundary condition $\Psi(0)=0$ if we neglect the contribution of the spin orbit coupling in the spinors. If so, the only reasonable wave function is obtained by
\begin{eqnarray}
\Psi(x)=\frac{1}{\sqrt{N}}(-1+i,i-1,-i-1,-1-i)^T\nonumber\\
\sin\left(k_{F} x\right)\exp\left[-\int_0^x dx' \frac{a_{\Delta}(\alpha,\mu,B)\Delta(x')}{\nu}\right].
\label{Eq:sin}
\end{eqnarray}

\subsection{Overlap of wavefunctions}

The analysis of Secs. \ref{Sec:subC}-\ref{Sec:subE} is done for isolated Majorana fermions in (semi-)infinite space with a spatial variation of the superconducting pairing. However, since Majorana fermions always appear in pairs and our system is finite, there can be a finite hybridization energy between them. 
The hybridization energy is directly related to the overlap of the two wavefunctions. For $\Delta(x)$ defined in Eq.~(\ref{pairing}), in the regime where $B<B_{c}$, we can approximate the solution at the left end of the wire, where the proximity induced pairing decreases to zero, by the wavefunction of Eqs.~(\ref{Eq:MajoranaSol}),~(\ref{Eq:sin}), when spin-orbit coupling is strong/weak. On the other hand, around $x=x_B$ we have to take into account the wavefunction of Eq.~(\ref{Eq:Chi1}). For the strong spin orbit coupling regime, Eq.~(\ref{Eq:MajoranaSol}) has an oscillatory and a non-oscillatory part, while Eq.~(\ref{Eq:Chi1}) is non-oscillatory. Therefore, the hybridization energy will also be constituted by an oscillatory and a non-oscillatory part. Which of them is dominant is strongly dependent on the corresponding decay length and the relative position of the states.
The hybridization energy is expected to show an oscillatory nature if the oscillatory part of Eq.~(\ref{Eq:MajoranaSol}) is dominant when $x=x_B$ is approached, This is the case if
\begin{equation}
\int_0^{x_B} dx (1+a_\Delta (\alpha,\mu,B))\Delta(x)- \sqrt{B^2-\mu^2}< 0.
\label{Eq:Relation4}
\end{equation}
For $\Delta(x)$, following Eq.~(\ref{pairing}) with $B^2<\Delta_0^2+\mu^2$, $x_B$ is determined by
\begin{equation}
x_B=\xi \mathrm{arctanh}\left(\frac{\sqrt{B^2-\mu^2}}{\Delta_0}\right).
\label{Eq:xB}
\end{equation}
\begin{figure}
\includegraphics[width=1\linewidth]{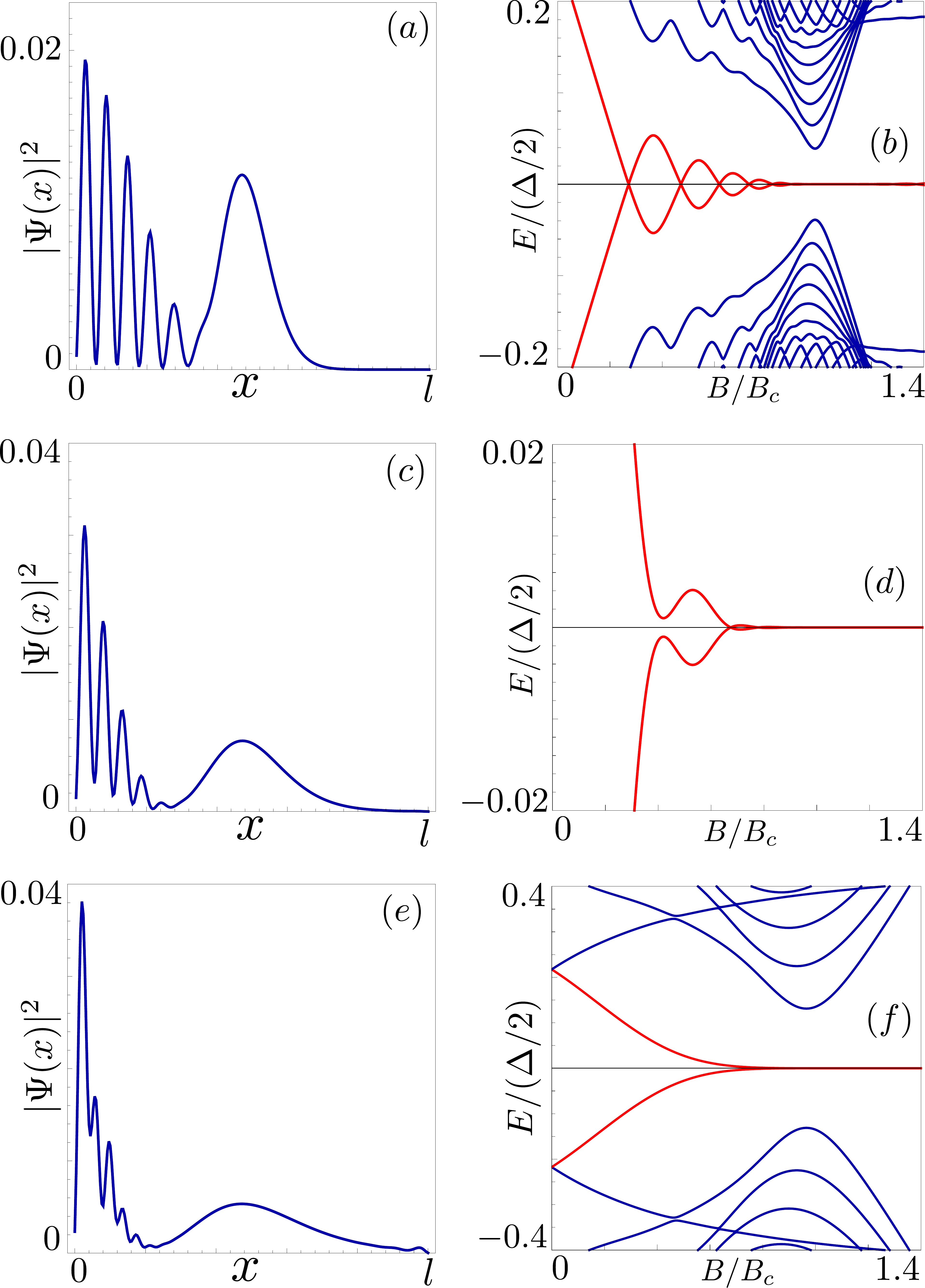}
\caption{Lowest energy eigenfunctions as a function of $x$ (left) with $B=1.134$ meV and eigenenergies (right) of the spin orbit coupled wire with proximity induced s-wave pairing with $\Delta_0=1.25$ meV, $N=250$, $L=2.5~\mathrm{\mu m}$, $\xi=0.8~\mathrm{\mu m}$, $\mu=0$ for three different values of the spin orbit coupling: (a)-(b) $\alpha=-15.12$ meVnm, (c)-(d) $\alpha=-37.8$ meVnm and (e)-(f) $\alpha=-75.6$ meVnm.}
\label{Fig:OscCombo}
\end{figure}
After performing the integration, we obtain
\begin{eqnarray}
R(\alpha,\mu,B)&\equiv &-\frac{1}{2}[1+a_\Delta(\alpha,\mu,B)]~\mathrm{ln}(1-\eta^2)\nonumber\\
&-&\eta~\mathrm{arctanh}(\eta)< 0,
\label{Eq:Relation5}
\end{eqnarray}
with $\eta=\sqrt{B^2-\mu^2}/\Delta_0$. $R(\alpha,\mu,B)$ is illustrated for different values of $a_{\Delta}(\alpha,\mu,B)$ in Fig. \ref{Fig:Relation}.
\begin{figure}
\includegraphics[width=1.05\linewidth]{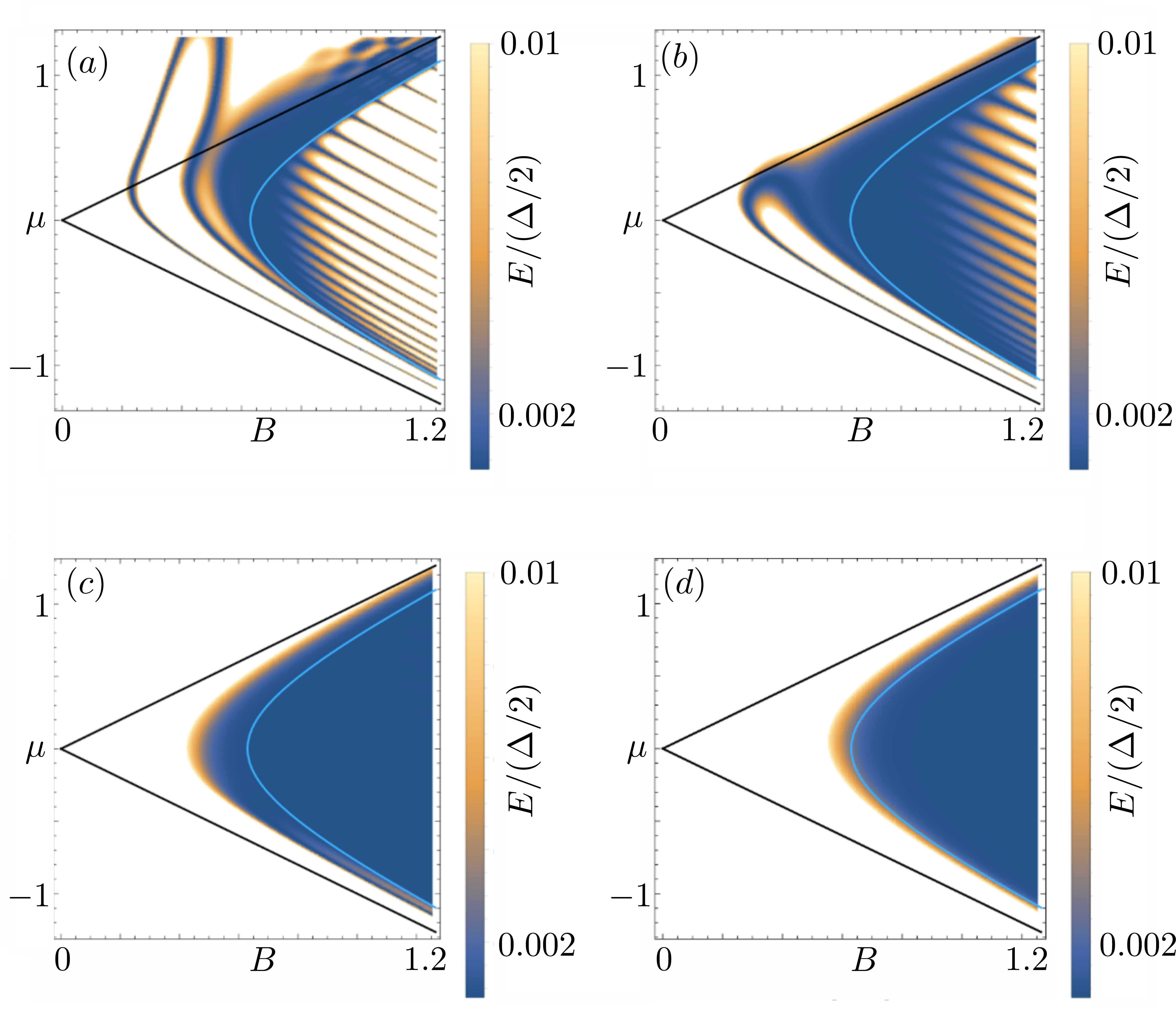}
	\caption{Numerical results for the lowest energy eigenvalue as a function of Zeeman energy $B$ and chemical potential $\mu$ in meV. The calculations are done for: $\Delta_0=0.63$ meV, $\xi=0.5~\mathrm{\mu m}$, $L=2~\mathrm{\mu m}$ and different spin-orbit coupling: (a) $\alpha=-10.1$ meVnm, (b) $\alpha=-20.2$ meVnm, (c) $\alpha=50.4$ meVnm and (d) $\alpha=-100.8$ meVnm.}
	\label{Fig:AlphaDep}
\end{figure}
If $a_{\Delta}(\alpha,B,\mu)\rightarrow 1$, which is the case in the strong spin-orbit regime, Eq.~(\ref{Eq:Relation5}) can not be fulfilled. Hence, the long wave contribution will always dominate the behavior of the wavefunction at $x=x_B$ and the hybridization energy will show a non-oscillatory behavior, which is coherent with numerical results (Fig. \ref{Fig:OscCombo} (e), (f)). For very large $\alpha$, similar arguments hold for the hybridization energy in the topological phase. Since the decay length of the wavefunction in the strong spin-orbit regime is proportional to $\alpha$, the overlap of the different wavefunctions is large, resulting in a suppression of zero-modes before the topological phase for large $\alpha$ (see Fig. \ref{Fig:AlphaDep}). \\

For $a_\Delta(\alpha,\mu,B)< 1$, on the other hand, the inequality Eq.~(\ref{Eq:Relation5}) can be fulfilled for some values of $B$ within $B^2<\Delta_0^2+\mu^2$ (see Fig. \ref{Fig:Relation} (b)).
The regime of dominant oscillatory behavior is amplified for small $a_\Delta(\alpha,\mu,B)$, i.e. weak spin-orbit coupling, a behavior explaining the numerical results (see Fig. \ref{Fig:OscCombo} (a), (b)).
In this regime, the wave function at $x=x_\nu$ provides strongly oscillatory character (Eq.~(\ref{Eq:sin})) leading to an oscillatory behavior of the hybridization energy (see Fig. \ref{Fig:OscCombo} (a)-(b)). \\
In the region between strong and weak spin-orbit coupling, instead, it is difficult to determine the analytical form of the wave function around $x_{\nu}$. It will be a mixture of Eqs.~(\ref{Eq:MajoranaSol})~and~(\ref{Eq:sin}). This is the regime, where we witness anticrossings in the hybridization energy, as the oscillatory and non-oscillatory contribution to the wave function at $x=x_B$ have similar decay lengths (Fig. \ref{Fig:OscCombo} (c)-(d)).\\

For smaller values of the gap parameter $\Delta_0$, all results remain qualitatively valid. However, the localization of the wavefunctions is reduced resulting in two major physical effects: Firstly, the overlap of the wavefunctions increases, and so does the hybridization energies. Secondly, the wavefunction centered around $x_B$ can significantly deviate from the Gaussian profile and can acquire oscillating contributions. This results in more complex hybridization energies close to the topological phase transition. The reason is that when the localization length of the wavefunction around $x_B$, which increases as $\Delta_0$ decreases, becomes comparable with the distance to the right end of the wire, then Friedel-like finite size oscillations become prominent \cite{Egger1995a, Fabrizio1995a, Eggert2009a, Kyl2016a}. Since the two length scales involved are the localization length of the wavefunction and the distance to the boundaries of the wire, these effects become more pronounced in short samples.
 
\subsection{Role of the chemical potential}

In the intermediate $\alpha$ regime, which ranges around $\alpha\sim 10-50$ meVnm, the chemical potential $\mu$ plays a crucial role since, especially as $\mu\rightarrow B$, $a_\Delta(\alpha,\mu,B)$ is a strongly asymmetric function with respect to $\mu\rightarrow -\mu$ (see Eq.~(\ref{Eq:aDelta}) and Fig. \ref{Fig:Relation} (a)). As $a_\Delta(\alpha,\mu,B)$ controls the gap size at $k=\pm k_F$, oscillations are more pronounced in the negative $\mu$ regime, which is indeed consistent with numerical results (see Fig. \ref{fig:e_B_mu} (d)). This allows us to witness low energy eigenvalues with oscillatory, anticrossing or monotonous convergence to zero also in dependence of the chemical potential $\mu$.

For experimentally relevant values of $\alpha\sim 20$ meVnm, we indeed expect to be in the transition regime between strong and weak spin-orbit coupling. Interestingly, the tendency to an asymmetric behavior with respect to $\mu$ is maintained qualitatively even for experimantally relevant coherence length.

With respect to our findings, the signature of a low energy conductance measurement could give an indication to the magnitude of the spin-orbit coupling as well as the chemical potential inside the wire.   

\section{Conclusion}
\label{s5}
Majorana fermions, i.e. zero-energy bound states, in a spin-orbit coupled quantum wire can exist even when the wire is not in the topological regime. The requirement is a finite coherence length $\xi$ of the proximity induced superconducting pairing amplitude. For slowly varying $\Delta(x)$ (large coherence length), we have given analytical and numerical demonstrations that the existence of zero-modes is possible in the whole $B\geq |\mu|$ region. Moreover, we have demonstrated that the momentum-space decomposition of the two Majorana fermions that can form before the topological phase transition are profoundly different. The one located at the end of the wire has an oscillating wavefunction, while the one located at the end of the locally topological region of the wire has a non-oscillating structure. This particular behavior implies a rich scenario for the hybridization energy. As a function of spin-orbit coupling and chemical potential, different behaviors can be obtained, ranging from a pattern of decaying oscillations, to anticrossings, and to a monotonous decay to zero energy. Oscillations are favored by weak spin-orbit coupling and tendentially small chemical potential, and their amplitude decays as a function of the magnetic field because the two Majorana fermions get separated in space. Stronger spin-orbit coupling and higher chemical potential favor, on the other hand, anticrossings and monotonous decay. We have interpreted the results by means of effective Dirac-like models, which allowed us to understand them as a consequence of different decay lengths characterizing the various momentum components of the Majorana fermion wavefunctions.

\begin{acknowledgments}We thank L. Chirolli for interesting discussions. Financial support by the DFG (SPP1666 and SFB1170 "ToCoTronics"), the Helmholtz Foundation (VITI), and the ENB Graduate school on "Topological Insulators" is acknowledged. 

\end{acknowledgments}


%

\end{document}